\documentclass{PoS}

\newcommand{\tr}{{\rm Tr~}}

\def\lsim{\raise0.3ex\hbox{$<$\kern-0.75em\raise-1.1ex\hbox{$\sim$}}}
\def\gsim{\raise0.3ex\hbox{$>$\kern-0.75em\raise-1.1ex\hbox{$\sim$}}}
\def\bx{{\bf x}}

\newcommand{\be}{\begin{enumerate}}
\newcommand{\ee}{\end{enumerate}}
\newcommand{\bi}{\begin{itemize}}
\newcommand{\ei}{\end{itemze}}
\newcommand{\beq}{\begin{equation}}
\newcommand{\eeq}{\end{equation}}
\newcommand{\beqn}{\begin{eqnarray}}
\newcommand{\eeqn}{\end{eqnarray}}
\newcommand{\lla}{\left\langle}
\newcommand{\rla}{\right\rangle}

\title{Renormalized Polyakov loops in various representations in finite temperature SU(2) gauge theory}

\ShortTitle{Renormalized Polyakov loops in SU(2)}

\author{\speaker{Kay H\"ubner}\\%\thanks{A footnote may follow.}\\
        Brookhaven National Lab\\
        Upton, NY 11973\\
        USA\\
        E-mail: \email{huebner@quark.phy.bnl.gov}}

\author{Claudio Pica\\
        Brookhaven National Lab\\
         Upton, NY 11973\\
         USA\\
        E-mail: \email{pica@quark.phy.bnl.gov}}

\abstract{
We present results for the renormalized Polyakov loop in the three lowest irreducible representations of SU(2) gauge theory at finite temperature. We will discuss their scaling behavior near $T_c$ and test Casimir scaling in the deconfined phase. Moreover, we will compare these results to calculations for the renormalized Polyakov  loops in several representations in the SU(3) gauge theory.
}

\FullConference{The XXVI International Symposium on Lattice Field Theory\\
		 July 14-19 2008\\
		 Williamsburg, Virginia, USA}

\begin{document}

\section{Introduction}

The Polyakov loop is the order parameter of the deconfinement phase transition in SU($N_c$) gauge theories and needs renormalization when calculated on the lattice \cite{Polyakov:1980ca}. 
Approximate Casimir scaling of irreducible representations seems to be an important property of gauge theories.
Studies have found it to be valid in potentials at $T=0$ in SU($N_c$) \cite{Heller:1984hx,Schroder:1998vy,Piccioni:2005un,Bali:2000un,Bicudo:2007xp} and even in the exotic group $G_2$ \cite{Liptak:2008gx}. 
Free energies and the Polyakov loop in different representations in the deconfined phase of SU(3) also exhibit this property \cite{Doring:2007uh,Hubner:2007qh,Gupta:2007ax}.
As an application, recent studies calculating the shear vicosity in a large-$N_c$ setting also use renormalized Polyakov loops \cite{Hidaka:2008dr}. 
In this work we compute renormalized Polyakov loops in the three lowest irreducible representations of SU(2).

\section{Operators and Lattice Implementation}

In SU(2) the character of the fundamental representation on the lattice is given by 
\beq
   \chi_1({\bf x}) = \tr \prod_{x_0=0}^{N_\tau-1} U_4({\bf x},x_0).
\eeq
Characters in higher representation can be obtained through the recursion formula
\beq
\chi_{n+1} = \chi_n \chi_1 - \chi_{n-1},\quad \chi_0=1,
\eeq
where $n\in\mathcal{N}$ and $j=\frac{n}{2}$ is the isospin of the representation with dimension $D(n)=2j+1$ and the eigenvalue of the quadratic Casimir operator is given by $C_2(n)=j(j+1)$.
This can be done at every spatial point $\bx$.
The ratio of Casimirs in representation $n$ and the fundamental representation is given by
\beq
d_n=\frac{C_2(n)}{C_2(1)}=\frac{4}{3}j(j+1).
\eeq
%  
%Table \ref{tab:rep} summarizes some important quantities for the three lowest irreducable representation relevant for this work.
The bare Polyakov loop in representation $n$ is then defined by
\beq
L^b_n=\frac{\sum_{\bf x} \chi_n({\bf x})}{D(n)V}.
\label{eq:bare_loop}
\eeq

We have calculated bare Polyakov loops (\ref{eq:bare_loop}) in the three lowest irreducible representations on lattices using the standard Wilson action with lattice sizes $N_\tau=4,5,6,8,12$ and $N_\sigma/N_\tau=4$ and $N_\sigma/N_\tau=8$ for some couplings to check for finite volume effects.
Statistics where as high as $O(10^5)$ close to $T_c$.

%
%\begin{table}
%  \begin{center}
%    \begin{tabular}{cccccl}
%      &&&&&\\
%      $j=n/2$ & $D(n)$ & $N_c$-ality & $C_2(n)$&$d_n$ & \\
%      \hline
%      $1/2$   & $2$ & $1$ &$3/4$& $1$ & fundamental \\
%      $1$   & $3$ & $0$ &$2$& $8/3$ & adjoint \\
%      $3/2$   & $4$ & $1$ &$15/4$& $5$ &  \\
%      \hline
%      &&&&&\\
%    \end{tabular}
%\caption{\label{tab:rep}Some quantities for the three lowest irreducable representation of SU(2).
%}
%  \end{center}
%\end{table}
%

\section{Renormalization Procedure}

Polyakov loops with smooth contours can be renormalized through a charge renormalization \cite{Polyakov:1980ca}. Thus the renormalized Polyakov loop in representation $n$ can be obtained by
\beq
   L_n(T)=Z_n(g^2)^{{{d_n}}N_\tau}\lla L^b_n\rla (g^2,N_\tau),
\label{eq:renorm_pol}
\eeq
where the renormalization constants $Z_n(g^2)$ only depend on the bare coupling and the representation $n$, $\lla\cdot\rla$ denotes the thermal average and $T=\left(a(g^2)N_\tau\right)^{-1}$ is the temperature on the lattice. Note that we have included a factor $d_n$ in definition of the exponent in (\ref{eq:renorm_pol}) for the $Z_n(g^2)$. 
We employ here a minor variant of a renormalization procedure for the Polyakov loop in arbitrary representations developed and applied to SU(3) in \cite{Gupta:2007ax}, to which we refer the reader for details.
This procedure uses bare loops obtained from at least two lattices with different temporal extend $N_\tau$ and the value of a renormalized Polyakov loop at one temperature as a seed value. The output are renormalized Polyakov loops in the deconfined phase (for $N_c$-ality zero representations also in the confined phase) and the corresponding renormalization constants. Since no values for renormalized Polyakov loops are known for SU(2), we use values for the renormalization constants as seed values instead, which can be obtained from the self-energy part $V_0$ of static potentials at $T=0$. 
The renormalization constants for the fundamental representation are then connected to $V_0$ by $Z_1(g^2)=\exp(V_0/2)$. 

Given the renormalization constant $Z_1$ at some coupling $g_{\mbox{seed}}^2$, the values of $L_n(g^2)$ and $Z_n(g^2)$ at different couplings $g^2$ are determined by our procedure. The number of points where $L_n$ and $Z_n$ are determined is a function of the initial seed coupling and of the two $N_\tau$ used. This number turns out to be small with our parameters so we made a linear interpolation between the two highest $\beta$'s listed in table \ref{tab:Renorm_const} in order to generate more seed values.
%Since the temperatures and couplings where the $L_n(T)$ and $Z_n(g^2)$ are obtained are fixed by the starting coupling and the $N_\tau$'s used and comes out to be small, we use a linear interpolation for $Z_1(g^2)$ for the two highest $\beta$'s listed in table \ref{tab:Renorm_const} in order to generate more seed values.
We will show later that this is consistent.
For higher representations we assume Casimir scaling (CS) $Z_n(g^2)=Z_1(g^2)$ for the seed values. 

\begin{table}
  \begin{center}
    \begin{tabular}{llll}
      &&&\\
      $\beta=4/g^2$ & $V_0$ & $Z_1$ & Ref\\
      \hline
      $2.5115$   & $0.537(4)$ & $1.308(26)$ & \cite{Bali:1993tz}\\
      $2.74$   & $0.482(3)$ & $1.2725(19)$  & \cite{Bali:1993tz}\\
      $2.96$   & $0.4334(9)$ & $1.2419(6)$  & \cite{Bali:1996cj}\\
      \hline
      &&\\
    \end{tabular}
    \caption{\label{tab:Renorm_const}Renormalization constants from SU(2) static potentials.
}
  \end{center}
\end{table}

\section{Results}

\begin{figure}
\scalebox{0.6}{\includegraphics{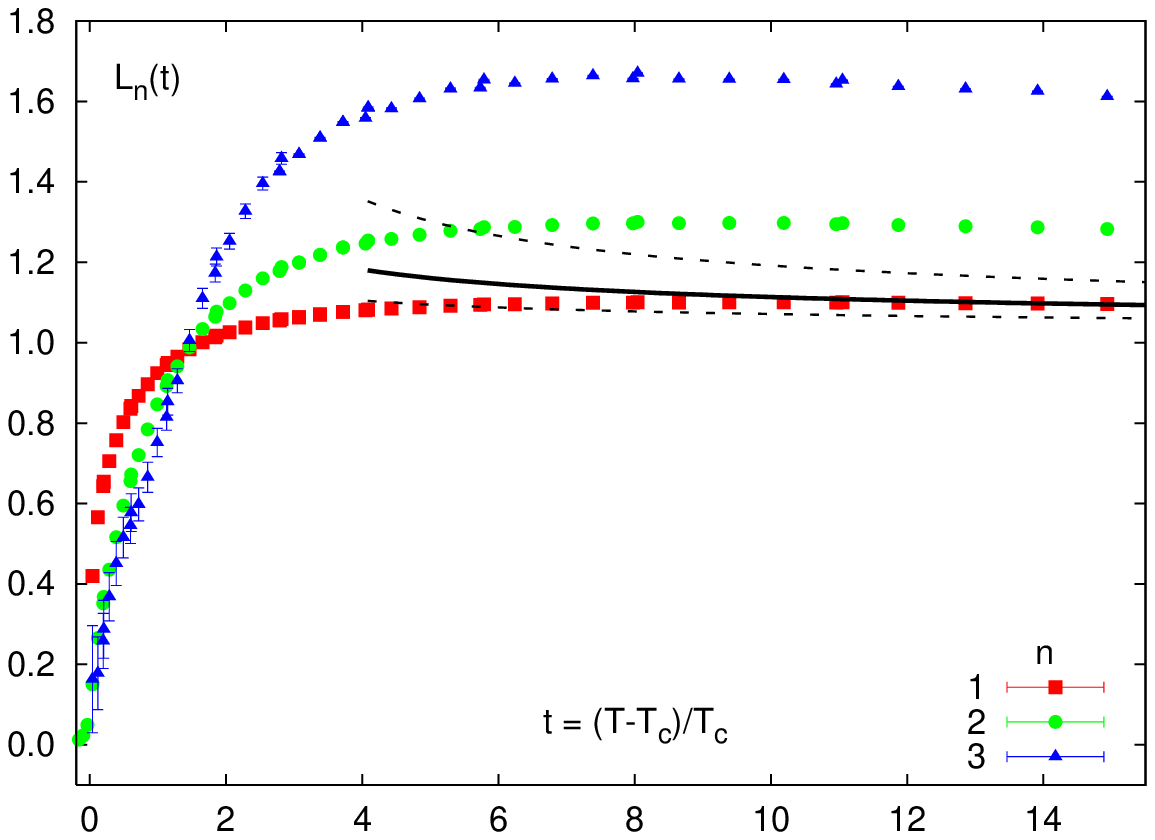}}
\scalebox{0.6}{\includegraphics{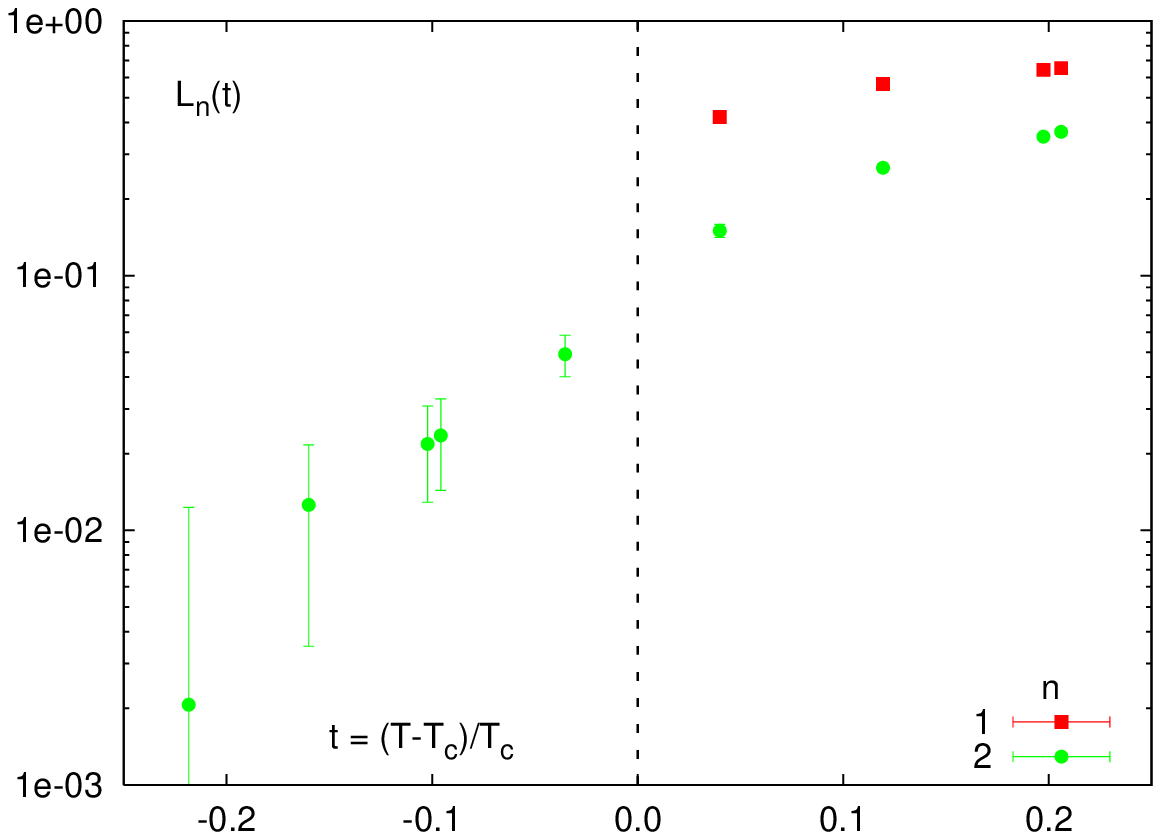}}
\caption{\label{fig:pol}Left: Renormalized Polyakov loops for $n=1,2,3$ in the deconfined phase. The black lines show a HTL-result \cite{Gava:1981qd} for $\mu=\pi/2,\pi,2\pi$ (from top to bottom). 
Right: The phase transition region. $L_2(T)$ is non-zero also in the confined phase. 
}
\end{figure}

Applying the renormalization procedure outlined above, we find a remnant $N_\tau$-dependence in the  renormalized Polyakov loops when using bare loops from $N_\tau<6$ lattices. Therefore we only present results from renormalization of bare loops from $N_\tau=6,8$ lattices in this work.
In fig.~\ref{fig:pol}(left) we show the renormalized Polyakov loops for $n=1,2,3$. The black lines are a HTL-results \cite{Gava:1981qd} for the fundamental Polakov loop for $\mu=\pi/2,\pi,2\pi$ (from top to bottom) using $T_c/\Lambda_{\overline{MS}}=1.23(11)$ \cite{Fingberg:1992ju} and the two-loop running coupling \cite{Bali:1992ru,Karsch:2000ps,Gupta:2000hr}. In the range  $[1-1.05]T_c$ some volume dependence might still persist. 
Fig.~\ref{fig:pol}(right) shows the phase transition region. $L_2(T)$ is clearly non-zero also in the confined phase, though a remnant volume dependence is still present.
These results are qualitatively very similar to the findings in SU(3) \cite{Gupta:2007ax} except close to $T_c$ of course, where the spin-half representations in SU(2) have to vanish. 

\begin{figure}
\scalebox{0.6}{\includegraphics{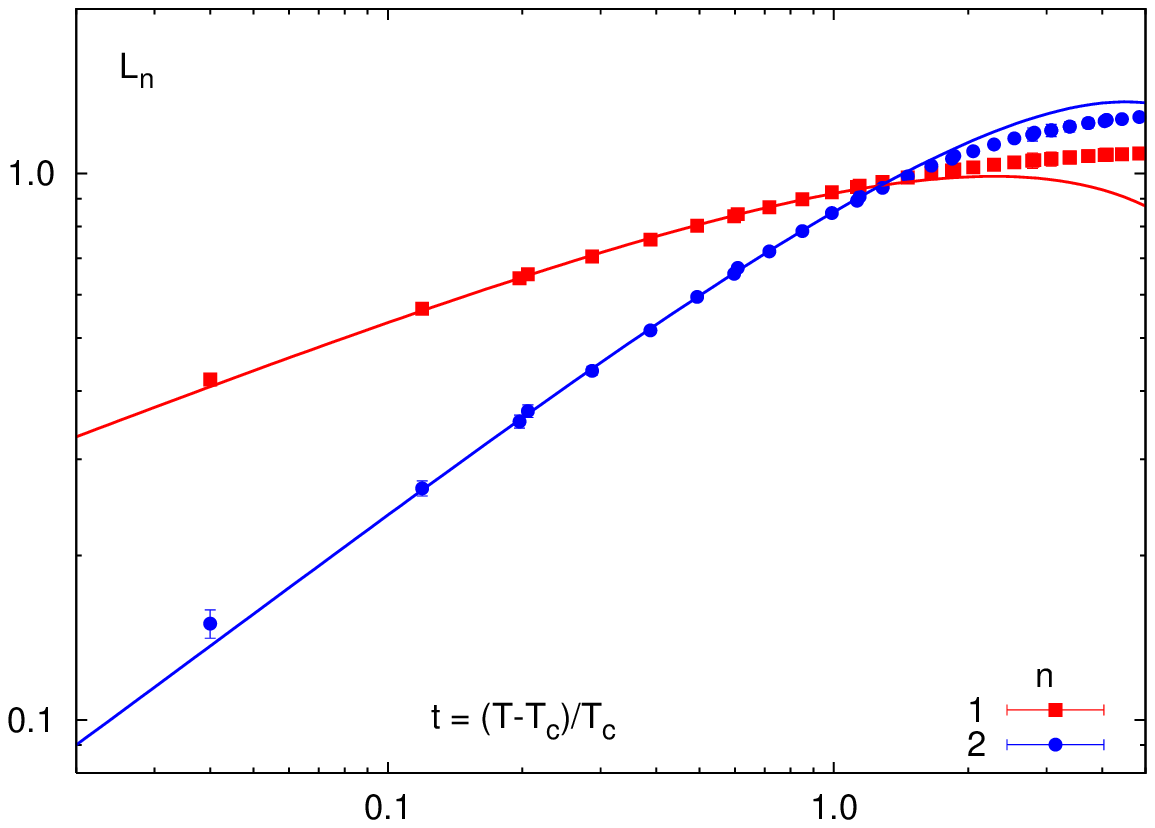}}
\scalebox{0.6}{\includegraphics{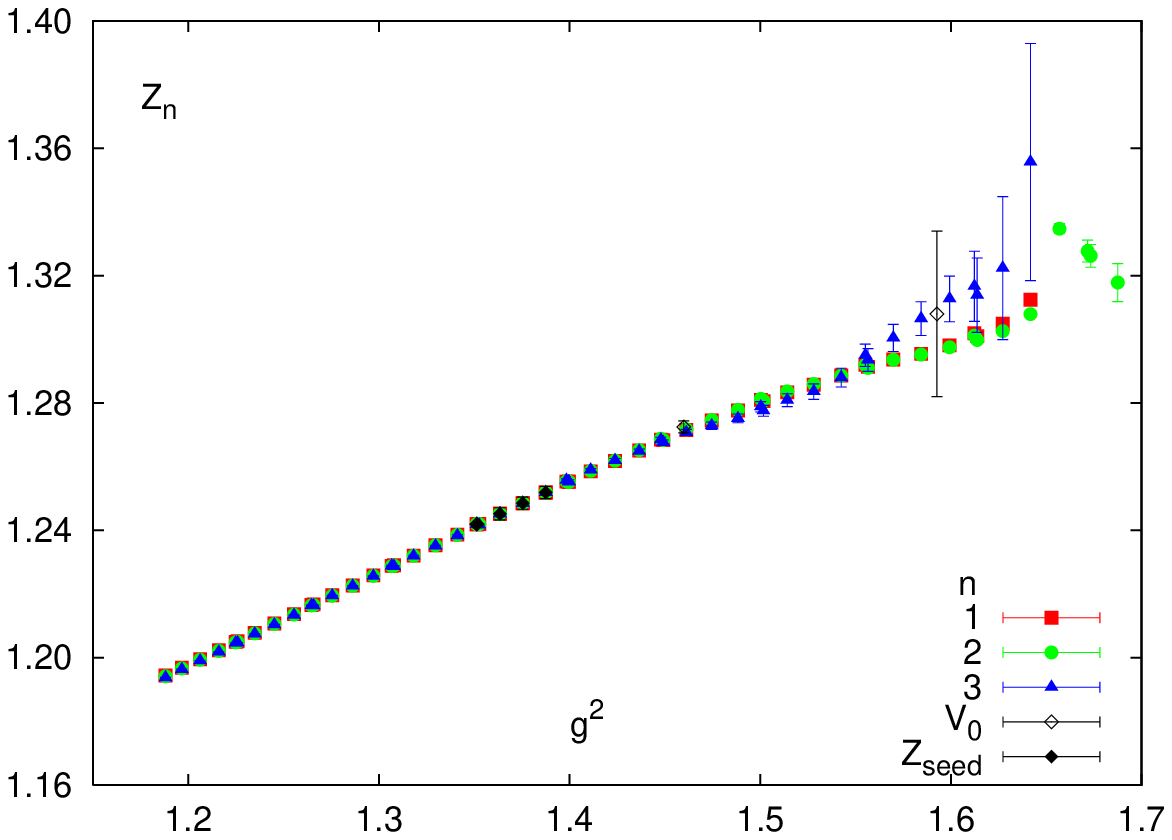}}
\caption{\label{fig:Zren}Left: Next-to-leading order fit for $L_1$ and $L_2$ with ansatz (4.2) and fixed exponents. Right: Renormalization constants. See text. 
}
\end{figure}

The fundamental Polyakov loop in SU(2) is expected to show critical scaling close to $T_c$.
For higher representations we make a phenomenological ansatz
%Moreover, a scaling behavior is also expected for higher representations, in leading order
%
\beq
L_n = A_n t^{n\beta} + c_n\delta_{d,0},\quad\mbox{where}\quad d:\mbox{ $N_c$-ality},
\label{eq:fit_lo}
\eeq
where $t=(T-T_c)/T_c$ is the reduced temperature.
Similar ans\"atze have been used on bare Polyakov loops \cite{Damgaard:1987wh,Redlich:1988db,Kiskis:1989hq}. In next-to-leading order we use
\beq
L_n =A_n t^{n\beta} (1+B_nt^{\Delta}) + c_n\delta_{d,0},
\label{eq:fit_nlo}
\eeq
where for the 3d-Ising universality class, in which SU(2) is in, we have $\beta=0.3265(3)$, $\omega=0.84(4)$, $\nu=0.6301(1)$ and $\Delta=\omega\nu=0.530(16)$ \cite{Pelissetto:2000ek}.
We have performed fits to $L_1$ and $L_2$ close to $T_c$ with ansatz (\ref{eq:fit_lo}) with fitting parameters $A_n,\beta$ and $c_2$ and with ansatz (\ref{eq:fit_nlo}) using exponents $\beta$ and $\Delta$ fixed to those of the 3d-Ising universality class and free parameters $A_n,B_n$ and $c_2$. Table \ref{tab:fits} shows the resulting fit parameters, fig.~\ref{fig:Zren}(left) the nlo-fit.  
The leading order fit returns a smaller $\beta$ than expected from universality, which is explained by the negative $B_n$ in the nlo-fit, which describes the data well up to ca.~$2T_c$. The value of $L_2(T_c)$ is badly constraint and compatable with zero. 

In fig.~\ref{fig:Zren}(right) we show the renormalization constants obtained from  $T=0$ self-energies ($V_0$), our interpolated seeds ($Z_{\mbox{seed}}$) and the results of the renormalization procedure for $Z_n$ and $n=1,2,3$. All $Z_n$ coincide very well up to $g^2\lsim 1.55$, where $Z_3$ breaks away and becomes noisy.
Moreover, the $Z_n$ are consistent with the seed data and those from $V_0$ calculations. Thus $Z_n(g^2)=Z_1(g^2)$ for almost all couplings used means that violations of CS by the UV-divergencies are small. The same behavior has been found in SU(3) \cite{Gupta:2007ax}.

%
%
%\begin{figure}
%\scalebox{0.6}{\includegraphics{fit_lo.eps}}
%\scalebox{0.6}{\includegraphics{fit_nlo.eps}}
%\caption{\label{fig:fit}Left: leading order fit with free exponent. Right: next-to-leading order fit with fixed exponents. 
%}
%\end{figure}
%
%
\begin{table}
  \begin{center}
    \begin{tabular}{llllll}
      &&&&\\
      lo Eq (\ref{eq:fit_lo})&$n$ & Interval &$A_n$ & $n\beta$ & $c_n$\\
      \hline
      &$1$   & $[0.01:0.25]$  & $0.9952(89)$ & $0.2666(46)$ & -\\
      &$2$   & $[0.1:0.5]$  & $0.881(24)$  & $0.576(56)$  & $0.007(40)$ \\
      \hline
      \hline
      nlo Eq (\ref{eq:fit_nlo}) &$n$ & Interval &$A_n$ & $B_n$ & $c_n$\\
      \hline
      &$1$   & $[0.01:1.0]$ & $1.2203(69)$ & $-0.2464(65)$ & -\\
      &$2$   & $[0.1:1.0]$ & $1.126(40)$ & $-0.249(20)$ &  $0.005(10)$ \\
      \hline
      &&&&\\
    \end{tabular}

\caption{\label{tab:fits}Fit results for $L_1$ and $L_2$ using leading order (lo) (4.1) with free exponent and next-to-leading order (nlo) (4.2) with fixed exponents.
}

  \end{center}
\end{table}

\begin{figure}
\scalebox{0.6}{\includegraphics{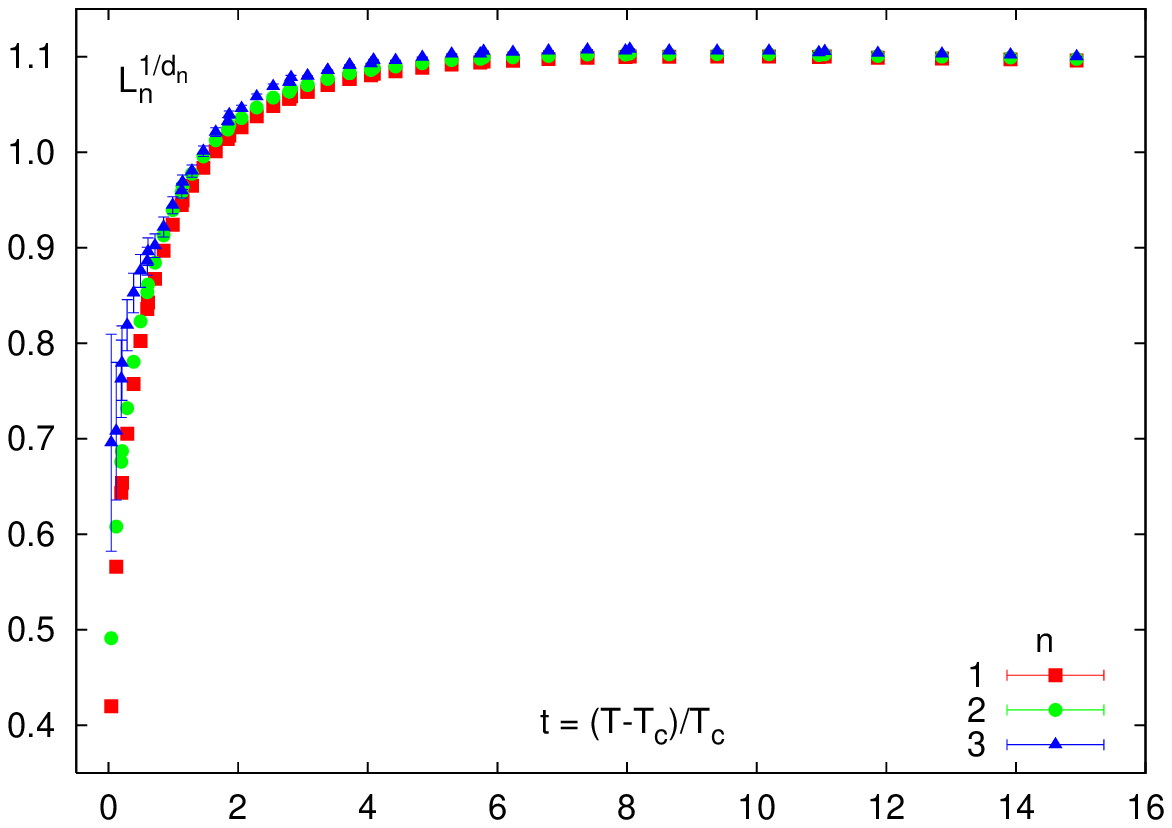}}
\scalebox{0.6}{\includegraphics{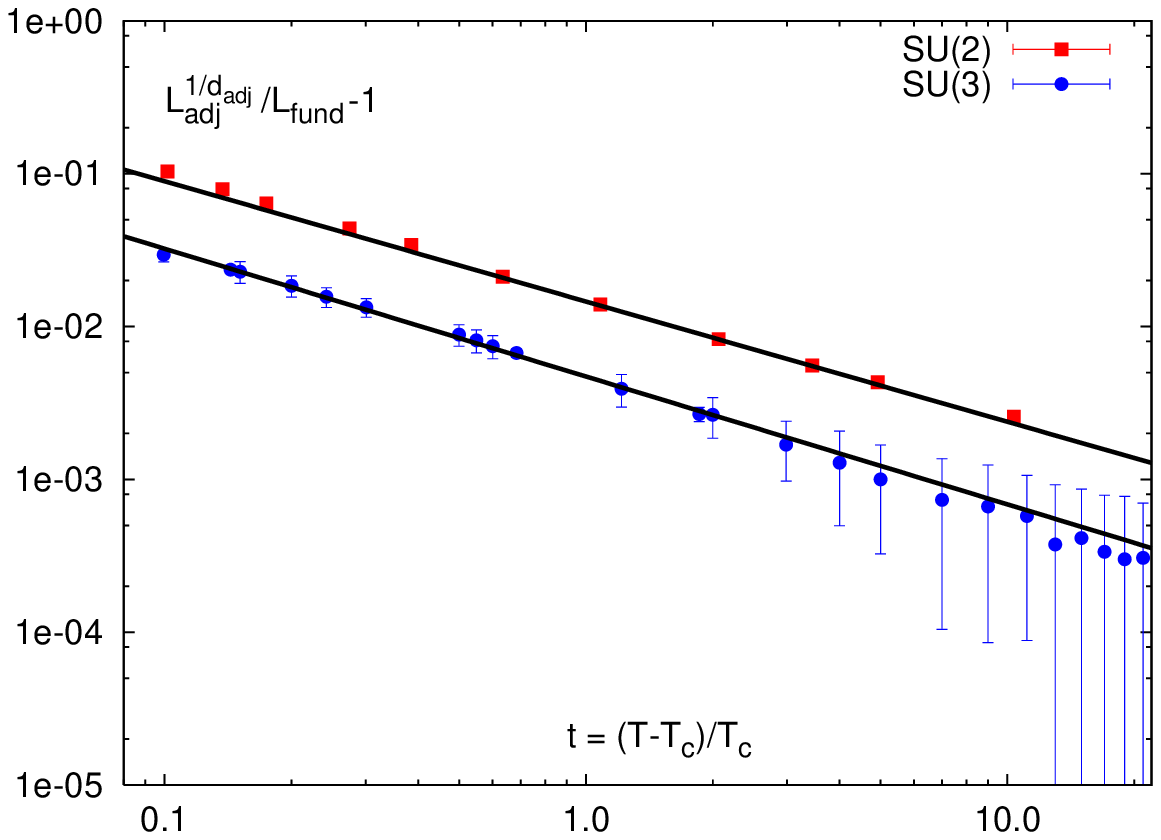}}
\caption{\label{fig:Casimir}Left: Scaled Polyakov loops. Right: Deviation from Casimir scaling for adjoint/fundamental loops in SU(2) and SU(3).
}
\end{figure}

We now check whether the renormalized Polyakov loops satisfy CS, i.~e.~$L_n(T)=\left(L_1(T)\right)^{d_n}$, which is expected to hold for large $T$ \cite{Schroder:1998vy}.
If this is correct, $\left(L_n(T)\right)^{1/d_n}$ is independend of $n$. Fig.~\ref{fig:Casimir}(left) shows, that CS is fullfilled very well at high temperatures. Close to $T_c$ deviations occur as expected.
More quantitative results can be obtained by looking at 
\beq
\frac{L_n^{1/d_n}}{L_1}-1,
\label{eq:dev_CS}
\eeq
which is zero if CS is valid. Since the renormalization constants are independend of $n$ as shown above, (\ref{eq:dev_CS}) is renormalization group independent. We therefore show in fig.~\ref{fig:Casimir}(right) the corresponding expression for the bare Polyakov loops from $N_\tau=4$ lattices because they are less noisy. They agree within errors with the corresponding expression using renormalized Polyakov loops in SU(2).
Moreover, we reanalyze data obtained in SU(3) \cite{Gupta:2007ax}. We observe a power law behavior over a large temperature interval stretching from $1.1T_c$ to $11T_c$. A fit with the ansatz $At^x$ gives $x=-0.7870(64)$ for SU(2) and $x=-0.8366(91)$ for SU(3). Closer to $T_c$ the SU(3) data have to approach a constant of course, whereas in SU(2) it should diverge.

\begin{figure}
\scalebox{0.6}{\includegraphics{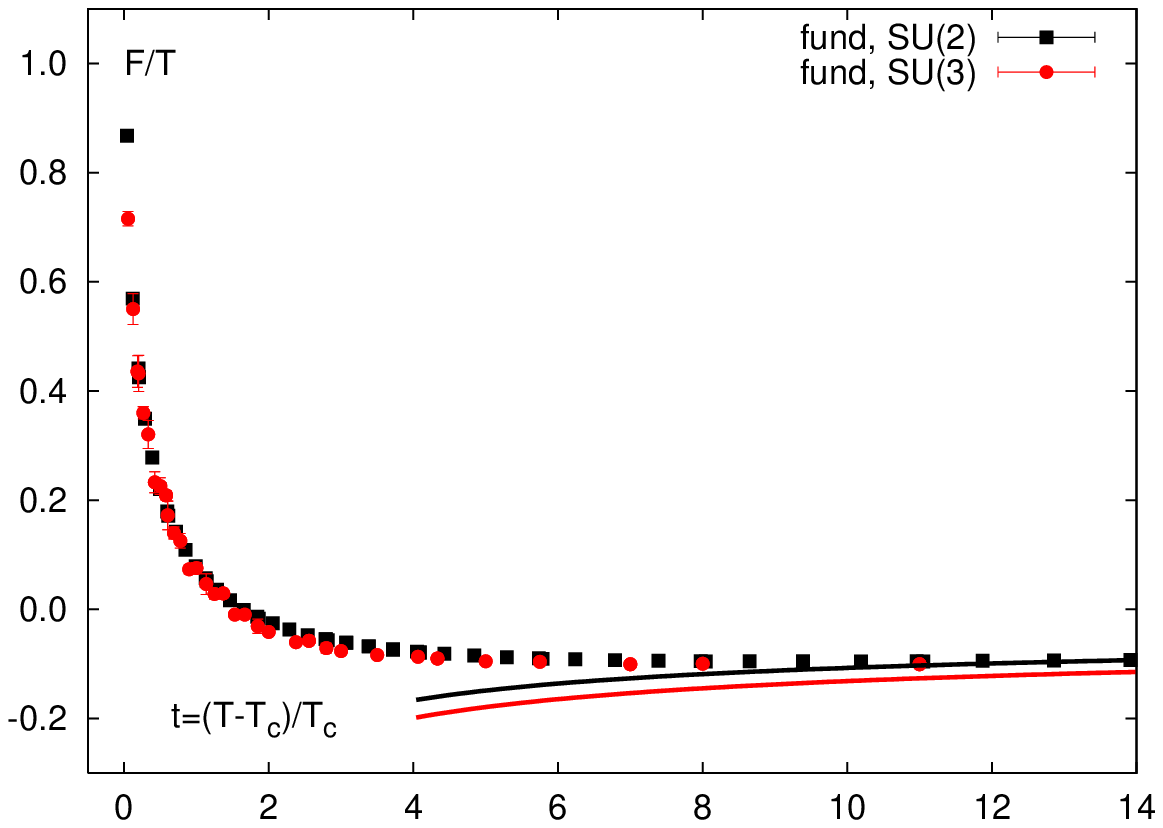}}
\scalebox{0.6}{\includegraphics{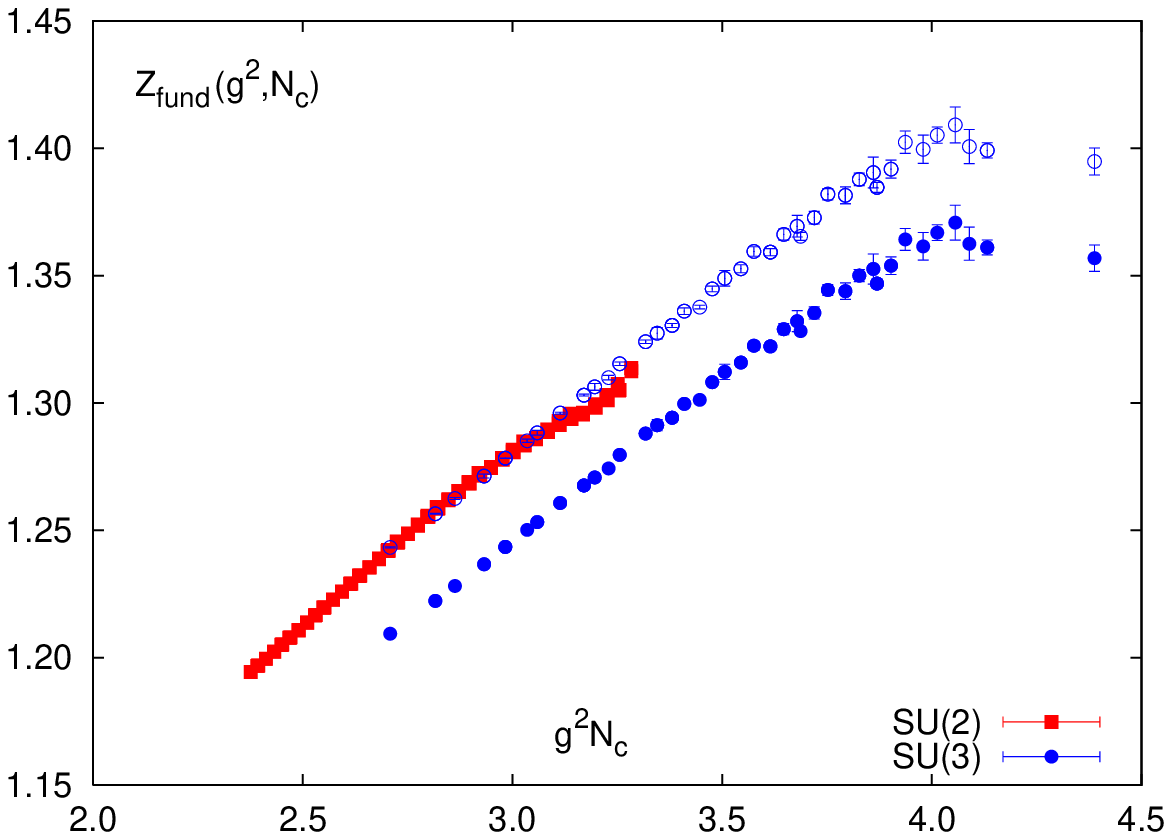}}
\caption{\label{fig:comparison}Left: free energies $F=-T\ln L$ belonging to the fundamental renormalized Polyakov loops in SU(2) and SU(3). Lines show the corresponding HTL-result \cite{Gava:1981qd}. Right: renormalization constants for the fundamental representation for SU(2) and SU(3) \cite{Gupta:2007ax} over the t'Hooft-coupling $g^2N_c$ (closed symbols). The open symbols show the SU(3)-data multiplied by $c=1.028$.
}
\end{figure}

We finally compare the free energies $F=-T\ln L$ belonging to the fundamental renormalized Polyakov loops in SU(2) and SU(3) in fig.~\ref{fig:comparison}(left). Close to $T_c$, $F(T,N_c=2)$ has to diverge, whereas $F(T,N_c=3)$ has to become constant. At large distances both free energies approach their corresponding HTL-result \cite{Gava:1981qd}. Nevertheless, both quantities show a remarkably similar behavior over almost the entire temperature interval studied here.    
In fig.~\ref{fig:comparison}(right) we plot the renormalization constants for the fundamental representation for SU(2) and SU(3) \cite{Gupta:2007ax} over the t'Hooft-coupling $g^2N_c$ (closed symbols). Open symbols show the SU(3)-data multiplied by $c=1.028$. 
The similarity of the free energies and 
the small multiplicative deviation in the renormalization constants might be explained by an already good $N_c$-scaling for this quantities, which has been shown to hold also for other quantities \cite{Lucini:2005vg}.

\section{Conclusions and Outlook}
We have computed the renormalized Polyakov loop in the three lowest irreducible representations in SU(2) gauge theory at finite temperature. We find the renormalization constants to be independent of representation. Casimir scaling for the renormalized Polyakov loops is realized at high temperatures in agreement with perturbative and other results. Deviations from Casimir scaling between the Polyakov loops in the fundamental and adjoint representations follow a power law over a very large temperature interval up close to $T_c$, which is also found to be the case in SU(3) reanalyzing data from \cite{Gupta:2007ax}. 
Comparing the renormalized fundamental Polyakov loops in SU(2) and SU(3) reveals a remarkable similarity over a large temperature interval in the deconfinement phase again up close to $T_c$. Renormalization constants in SU(2) and SU(3) show similar behavior with the t'Hooft-coupling with a small multiplicative deviations of $\approx 2.8\%$. 

In the future we want to extend our analysis on renormalized Polyakov loops in various representations at finite temperature to SU(4) and SU(6) gauge theories. This will allow us to check the large -$N_c$ scaling and Casimir scaling behavior of these quantities.

\section*{Acknowledgement}
%We like to thank the organisers of LATTICE 2008 for their hospitality.
We are grateful to Rob Pisarski, Frithjof Karsch, Sourendu Gupta and Olaf Kaczmarek for inspiring and fruitful discussions.
This manuscript has been authored under contract number DE-AC02-98CH10886 with the U.S. Department of Energy.

\end{document}